# Strong and tunable mode coupling in carbon nanotube resonators

Andres Castellanos-Gomez[*], Harold B. Meerwaldt, Warner J. Venstra, Herre S. J. van der Zant, Gary A. Steele[*]

Kavli Institute of Nanoscience, Delft University of Technology, Lorentzweg 1, 2628 CJ Delft, The Netherlands.

[*]E-mail: a.castellanosgomez@tudelft.nl, g.a.steele@tudelft.nl

The non-linear interaction between two mechanical resonances of the same freely suspended carbon nanotube resonator is studied. We find that in the Coulomb blockade regime, the non-linear modal interaction is dominated by single-electron-tunneling processes, and that the mode-coupling parameter can be tuned with the gate voltage, allowing both mode softening and mode stiffening behavior. This is in striking contrast to tension-induced mode coupling in strings, where the coupling parameter is positive and gives rise to a stiffening of the mode. The strength of the mode coupling in carbon nanotubes in the Coulomb blockade regime is observed to be six orders of magnitude larger than the mechanical mode coupling in micromechanical resonators.

Carbon nanotubes present remarkable properties for applications in nanoelectromechanical systems (NEMS) such as low mass density, high Young's modulus and high crystallinity [1,2]. This fact has motivated the use of carbon nanotubes to fabricate high-quality factor ($Q$) mechanical resonators [3] that can be operated at ultrahigh frequencies [4,5] and can be used as ultrasensitive mass sensors [6-8]. Additionally, both the mechanical tension and electrical properties of carbon nanotubes can be tuned to a large extent by an external electric field [9], making nanotubes a very versatile component in NEMS devices.

Due to the small diameter of carbon nanotubes, they can be easily excited in the non-linear oscillation regime [10]. Moreover, it has been demonstrated that the non-linear dynamics of carbon nanotubes can be tuned over a large range [11,12] making nanotube NEMS excellent candidates for the implementation of sensing schemes based on nonlinearity and for the study of fundamental problems on non-linear dynamics. The non-linear interaction between mechanical resonance modes is interesting both from a fundamental and from an applied perspective. Non-linear modal interactions have been studied recently in micro and nanoresonators [13-18]. These studies concentrated on mechanical coupling between the modes via the geometric nonlinearity or via the displacement-induced-tension, the same mechanism responsible for the Duffing non-linearity in doubly-clamped resonators. By employing a different mode of the same resonator as a phonon cavity, the mechanical mode can be controlled in-situ, and its damping characteristics can be modified to a great extent, leading to cooling of the mode and parametric mode splitting [13,16]. The non-linear coupling can be also used to detect resonance modes that would otherwise be inaccessible by the experiment [18] to increase the dynamic range of resonators by tuning the nonlinearity constant [18], and for mechanical frequency conversion [17]. Additionally, non-linear coupling has been proposed as a quantum non-demolition scheme to probe mechanical resonators in their quantum ground state [19] and as a way of generating entanglement between different mechanical modes [20]. Furthermore, recent theoretical work suggests that the interaction between mechanical resonances could be responsible for the spectral broadening in carbon nanotubes, thus limiting their *Q*-factor at room temperature [21].Despite the interest aroused by the modal interaction in nanotube resonators recently, experimental studies in this field are scarce.

Here we study the non-linear interaction between two different eigenmodes of a freely suspended carbon nanotube resonators at low temperatures, using a quantum dot embedded in the nanotube as a detector. We find that for nanotube resonators in the Coulomb blockade regime the non-linear modal interaction is dominated by single-electron-tunneling processes, as opposed to displacement-induced tension.. A strongly enhanced mode coupling is observed in the Coulomb-blockade regime, which is orders of magnitude larger than in conventional microresonators. Furthermore, in the Coulomb-blockade regime, the mode-coupling parameter can be tuned by adjusting the gate voltage, oscillating in sign over a gate range of only a few millivolts. This allows both mode softenening and mode stiffening behavior, in contrast to the case of tension-induced mechanical coupling in strings, where the coupling parameter is positive and gives rise to a stiffening of the mode.

The device consists of a single wall carbon nanotube suspended across a trench that bridges two metal electrodes (Fig. 1a). Electrons are confined in the nanotube by Schottky barriers at the metal contacts, forming a quantum dot in the suspended segment. The nanotube is grown in the last step of the fabrication process, yielding ultraclean devices which can have large quality factors. We perform all measurements in a dilution fridge at 20 mK. The carbon nanotube is actuated with a nearby antenna (separated about 2 cm from the sample). The detection of the resonator motion is carried out by monitoring the DC current while the nanotube is driven by the antenna. When the carbon nanotube is driven at resonance, its oscillation changes the capacitance between the nanotube and the gate, leading to an effective oscillating gate voltage which smears out the Coulomb peaks and thus yields a change in the DC current through the nanotube. More detail in this oscillation amplitude readout method, referred to here as the rectification method, can be found in Refs.[3, 12].

Figures 1b and 1c show two peaks in the DC current through the carbon nanotube that occur when the nearby antenna is driven at frequencies that match the mechanical resonances of the carbon nanotube (hereafter labeled mode A and mode B, respectively). The resonance frequencies of a clamped-clamped beam are calculated by Euler-Bernoulli theory as $f_n = \frac{\beta_n^2 r}{4\pi}\sqrt{E/\rho}$, with $\beta_n \cdot L$=4.73,7.85,11.00,etc. . Taking the Young's modulus $E$ = 1.3 TPa, the tube radius $r$ = 1 nm, the length of the suspended part of the tube $L$ = 600 nm and its mass density $\rho$ = 1400 kg/m$^3$, the resonance frequencies of the lowest three modes are $f_1$ = 151 MHz, $f_2$ = 415 MHz and $f_3$ = 814 MHz. These values correspond well with the measured resonance frequencies, $f_A$ = 175 MHz and $f_B$ = 957 MHz, indicating that the restoring force is dominated by the nanotube bending rigidity in this regime [22]. Note that even modes have a displacement that on average gives no change in capacitance during the oscillation. Because of this, the motion of even modes is not strongly excited by the antenna, nor is their motion strongly detected by the current through the nanotube. From a fit of the frequency response shown in figures 1(a) and 1(b), we extract a quality factor of $Q \sim 15\,000$ for the first mode and $Q \sim 5000$ for the second mode. This quality factor is lower than the highest values reported previously. We attribute the smaller $Q$-factor to damping from single-electron tunneling, as was reported in earlier devices [12]. In the device studied here, the quantum dot is less strongly tunnel-coupled to the source and drain, preventing us from measuring the mechanical resonance deep in the Coulomb valleys, where the previously reported high quality-factors were measured.

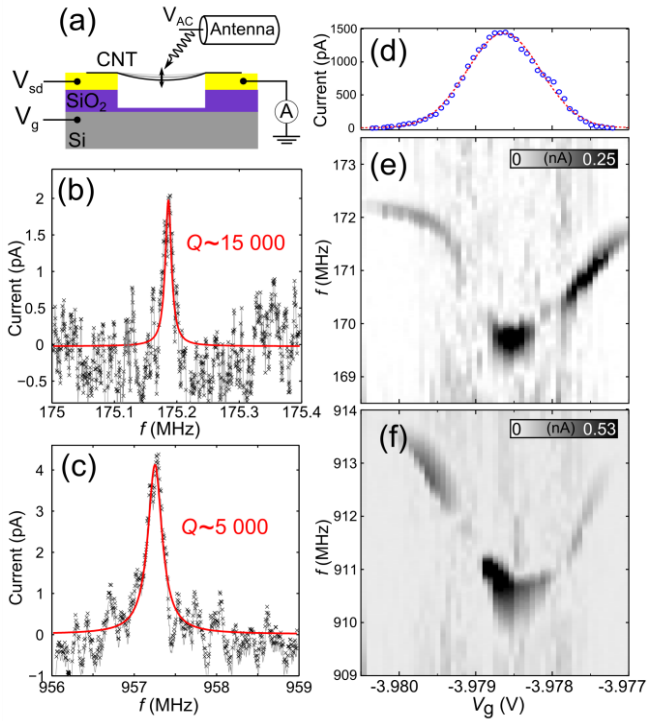

**Figure 1:** (a) Schematic diagram of the carbon nanotube device, freely suspended over a trench between drain and source electrodes. The Si substrate is employed as a backgate. (b) and (c) resonance spectra measured by means of the rectification method at fixed gate voltage and low excitation power for two mechanical eigenmodes of the carbon nanotube. The quality factor is obtained by fitting the measured spectra to the response of a dampened harmonic oscillator (solid red lines). (d) DC current through the nanotube *versus* gate voltage showing single-electron tunneling and Coulomb blockade electronic behaviour. (e) and (f) Color map showing the absolute value of the rectified current through the nanotube as a function of the RF frequency and gate voltage ($V_{sd} = 200\ \mu V$). The tuned mechanical resonance shows up as the gray/black curve with a dip at the Coulomb peak.

The mechanical origin of the resonance peaks observed in figure 1a and 1b is confirmed by studying the gate tunability of the resonance frequencies by modifying the tension in the nanotube, changing the applied gate voltage, which gives a tuning of 22 MHz/V and 38 MHz/V for the mode A and B respectively (not shown) [9]. Additionally, in the Coulomb blockade regime, the mechanical resonance frequencies experience a dip (down to 3MHz approx.) when the applied gate voltage is swept across a Coulomb peak (see Figures 1e and 1f) [12]. This is due to the electrostatic force on the nanotube, which depends on the average charge on the quantum dot. Across a Coulomb peak, the average charge increases monotonically from $N$ to $N+1$ electrons in a small gate voltage range [23-25]. When the nanotube is closer to the gate, the gate voltage is effectively larger, increasing the average charge on the carbon nanotube and causing a force towards the gate. This negative restoring force softens the carbon nanotube spring constant and results in a decrease of the mechanical resonance frequency.

To excite two mechanical resonances of the nanotube at the same time, the antenna is driven by the combined voltage of two RF signal generators. This allows one to study the interaction between mechanical modes in carbon nanotubes using a multifrequency experimental scheme, sweeping two frequencies at the same time. The frequency of the first signal generator is swept around the resonance frequency of mode A(fast axis sweep). Every time the frequency of the generator matches the resonance frequency of mode A, the DC current through the nanotube experiences a sudden change due to the above described rectification mechanism [3, 12]. After each sweep of the first generator, the frequency of the second generator is incremented and another sweep around the resonance of mode A is carried out with the first generator. This process is repeated until a sweep around the resonance frequency of mode B is accomplished (slow axis sweep). When the frequency of the second RF-generator is off-resonance with mode B, the oscillation amplitude of mode B is negligible and thus there is no appreciable coupling between modes A and B. On the other hand, when the frequency of the second generator approaches the resonance frequency of mode B, the oscillation amplitude of the carbon nanotube in mode B becomes appreciable and the resonance frequency of mode A may be modified by the motion of mode B. Mode B is driven at a much larger RF power (-5 dBm) than mode A (-20 dBm), to limit the backaction exerted by mode A on mode B.

From these measurements the DC current through the nanotube as a function of the two driving frequencies is obtained. The data can be conveniently represented in a (3-dimensional representation) color map form as shown in Figure 2, with a fast axis (frequency sweep around the resonance of mode A) and a slow axis (frequency sweep around the resonance of mode B). In the colorscale data, we have subtracted off the background signal from the mechanical resonance corresponding to the slow sweep direction. This background signal, which contains information about the response of mode B, is shown as a blue line profile overlayed on top of the colorscale data. The blue line profile indicates the horizontal position at which the frequency of the second generator hits the resonance of mode 2. These profiles show that the mechanical resonance of mode B presents a clear sharkfin shape, a characteristic of the Duffing-like nonlinear resonance, indicating that mode B vibrates in the non-linear regime.

Figure 2(a) demonstrates that the resonance frequency of mode A decreases when mode B is driven at resonance. This change in the resonance frequency of mode A while the oscillation amplitude of mode B increases, indicates the interaction between the two vibration modes of the carbon nanotube. Interestingly, the interaction changes qualitatively when adjusting the gate voltage: when this measurement is repeated for slightly different gate voltages, the effect of this modal interaction is dramatically different. In figure 2a to 2c we present three different situations: (a) the modal interaction reduces the resonance frequency of mode A (mode softening), (b) the modal interaction increases the resonance frequency of mode A (mode stiffening) and (c) the effect of the modal interaction is negligible (modal interaction suppression). A similar mode coupling behavior has been observed in two other carbon nanotube devices with slightly different geometries.

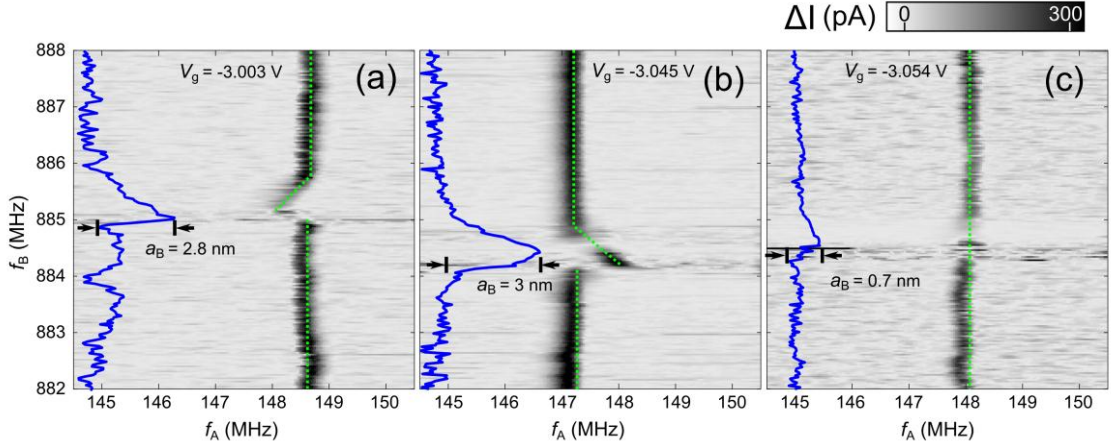

**Figure 2:** (color scale) DC rectified current through the nanotube as a function of the RF frequency of the two signal generators connected to the antenna ($V_{sd}$ = 200 µV). When the mechanical resonance of the mode A mode is excited, the current shows a sudden change. The resonance of mode A is marked by a drop in the rectified current (shown in black in the colormaps). (a) The resonance frequency of mode A shifts to lower frequency when the RF signal of the second generator hits the resonance frequency of the mode B ($V_g$ = -3.003 V). (b)-(c) same as (a) but with a $V_g$ value of -3.0045 and -3.0054 respectively, showing mode stiffening (b) and negligible coupling (c). The green dotted lines are a guide to the eye. The blue line profiles inserted in (a)-(c) show the oscillation amplitude of the mode B (calculated from the measured change in the DC current through the nanotube using an electrostatic model and assuming that the mode shape resembles that of the third bending mode of a doubly-clamped beam) as a function of the (slowly incremented) frequency of the second generator.

The (scaled) equation describing the motion of mode *i* taking into account the coupling between the modes can be written as

$$\ddot{u}_i + \eta_i \dot{u}_i + \omega_i^2 u_i + \sum_{j,k,l} \alpha_{ijkl} u_j u_k u_l = f_i \cos(\Omega_i t)$$

where $\eta_i$ denotes the damping, $\omega_i$ the resonance frequency and $f_i$ the driving force of mode *i*, and the indices *j,k,l* run over the number of modes considered. For a single mode (*i = j = k = l*) this yields the Duffing equation of a non-linear resonator. In a previous work in micromechanical resonators the *α* terms were shown to be due to displacement-induced tension [18]. For carbon nanotube quantum dots, however, it is established that the non-linear *α* terms, responsible of the Duffing non-linearity, are strongly influenced by the single-electron tunnelling processes in the suspended carbon nanotube [12]. This yields to interesting phenomena such as the gate tunability of the *α* terms in carbon nanotubes from positive to negative values in response to a small change in gate voltage.

To verify that the observed modal interaction is dominated by single-electron tunneling processes, we have measured the frequency tuning by the modal interactions as a function of the gate voltage. Figure 3a shows the maximum change in the resonance frequency of mode A (due to the interaction with mode B) at different gate voltages across a Coulomb peak, showing a continuous transition from stiffening to softening behaviour. The observed change in sign of the modal interaction (*i.e.* stiffening *vs.* softening behavior) is directly related to the sign of the non-linear spring constant α of the carbon nanotube, which can be obtained from the curvature of the $f_A$ *vs*. $V_g$ trace ($\partial^2 f_A/\partial V_g^2$) shown in figure 3b [12], indicating that the mechanism behind this gate dependence is the time-varying electrostatic force induced by the charge fluctuation in the quantum dot formed by the suspended carbon nanotube.

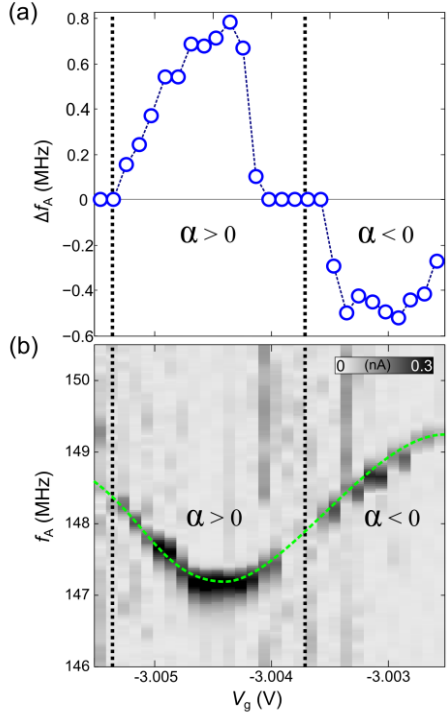

**Figure 3:** (a) maximum change in the resonance frequency of the mode A as a function of the gate voltage. The resonance frequency of mode A increases with gate voltages when the non-linear spring constant is positive (i.e. for the gate voltages at which Coulomb peaks occur). On the other hand, the resonance frequency of mode A decreases for the gate voltages when the non-linear spring constant is negative. (b) frequency dip of mode A when the gate voltage is swept across a Coulomb peak. The sign of the non-linear spring constant α can be determined by the sign of the curvature of the frequency *vs*. gate voltage trace.

Figure 4 shows the resonance frequency of mode A as a function of the oscillation amplitude of mode B. The resonance frequency shift shows a quadratic dependence with the oscillation amplitude of mode B, similar to the one observed in Ref. [18] for a micromechanical resonator. The modal interaction strength is given by the change in the resonance frequency as a function of the oscillation amplitude of mode B squared. The oscillation amplitude has been calculated from the measured change in the DC current through the nanotube using an electrostatic model and assuming that the mode shape resembles that of the third bending mode of a doubly-clamped beam. We find that the modal interaction strength can continuously be tuned from $63 \pm 8$ kHz/nm$^2$ (stiffening) to $-55 \pm 4$ kHz/nm$^2$ (softening). This modal interaction is not only highly tunable, with a fairly low change in the gate voltage, but it is also remarkably strong: six orders of magnitude larger that the coupling strength measured in micromechanical resonators (about 0.025Hz/nm$^2$) [18].

An interesting consequence of the strong mode coupling observed in suspended carbon nanotubes in the Coulomb blockade regime is that it could potentially provide an ultimate limit to the linewidth of the frequency response, and therefore also the quality factor of the resonator. At finite temperature, for example, the thermal motion of one mode will result in spectral broadening of the other mode. The third bending mode of the carbon nanotube device presented here at a temperature of 4.2 K would exhibit a Brownian motion of 50 pm, leading to a broadening of the mechanical linewidth of the fundamental mode of 170 Hz. Thus, the mode coupling observed here would limit the quality factor to a maximum value of $10^6$, although this maximum quality factor could be even lower if the interaction with other modes is taken into account. Such a spectral broadening is present even at zero temperature due to the zero point fluctuations: the expected zero point motion of 14 pm for the third eigenmode would yield a dispersive coupling of 12 Hz , and places a limit on the quality factor of $1.5 \cdot 10^7$. The observed quality factors in this device, however, are lower than these theoretical limits.

From a different point of view, the mode coupling observed here can be thought of as a non-linear coupling between two phonon cavities. In this case, the spectral broadening of 12 Hz calculated above using the zero-point fluctuations of the third mode would represent the strength of the dispersive coupling *g* between the two cavities in their ground state. For our device, the coupling rate is smaller than the decoherence rates of the modes ($\Gamma_{A,B}= f_{A,B} / Q_{A,B} \approx$ 10-200 kHz), placing the present experiment in the weak coupling limit. The strong coupling limit ($g > \Gamma_{A,B}$) could potentially be reached in devices with larger *Q* factors and sharper Coulomb peaks.

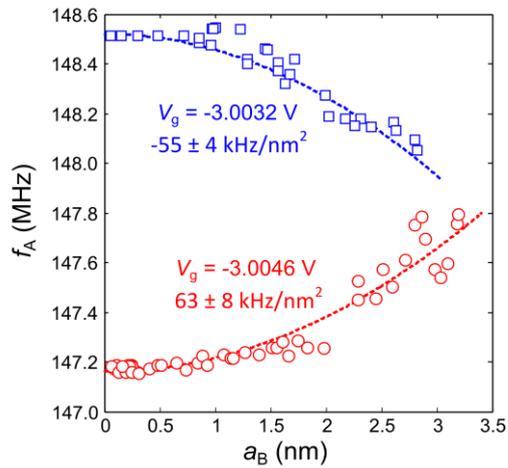

**Figure 4:** Resonance frequency shift of the mode A as a function of the oscillation amplitude of the mode B for two different gate voltages, showing softening (blue squares) and stiffening (red circles) behaviour. The modal coupling strength can be determined from the quadratic dependence of the frequency shift with the oscillation amplitude (dashed lines).

In conclusion, we have studied the non-linear interaction between two different eigenmodes in freely suspended carbon nanotube resonators. In the Coulomb blockade regime, the mode coupling in suspended carbon nanotubes is dominated by single-electron-tunneling processes. In contrast to purely mechanical mode coupling, in this regime both the strength and the sign of the coupling can be tuned by changing an external gate voltage. The modal interaction strength in carbon nanotubes is remarkably strong, about six orders of magnitude larger than that of previously studied micromechanical resonators.

**ACKNOWLEDGMENT**

This work was supported by FOM and the European Union (FP7) through the programs QNEMS and RODIN.